\documentclass[aps,prd,twocolumn,amsmath,amssymb,showpacs,showkeys]{revtex4}
\usepackage{bm}
\newtheorem{theorem}{Theorem}
\newtheorem{remark}{Remark}

\begin{document}


\title{The stationary Weyl equation and Cosserat elasticity}


\author{Olga Chervova}
\email[]{O.Chervova@ucl.ac.uk}
\affiliation{Institute of Origins,
University College London,
Gower Street,
London WC1E~6BT, UK}
\author{Dmitri Vassiliev}
\email[]{D.Vassiliev@ucl.ac.uk}
\homepage[]{http://www.ucl.ac.uk/Mathematics/staff/DV.htm}
\affiliation{Department of Mathematics and Institute of Origins,
University College London,
Gower Street,
London WC1E~6BT, UK}


\date{\today}

\begin{abstract}
The paper deals with the Weyl equation which is the massless Dirac
equation. We study the Weyl equation in the stationary setting, i.e.
when the spinor field oscillates harmonically in time. We
suggest a new geometric interpretation of the stationary Weyl
equation, one which does not require the use of spinors, Pauli
matrices or covariant differentiation. We think of our 3-dimensional
space as an elastic continuum and assume that material points of
this continuum can experience no displacements, only rotations. This
framework is a special case of the Cosserat theory of elasticity.
Rotations of material points of the space continuum are described
mathematically by attaching to each geometric point an orthonormal
basis which gives a field of orthonormal bases called the coframe.
As the dynamical variables (unknowns) of our theory we choose the
coframe and a density. We choose a particular potential energy which
is conformally invariant and then incorporate time into our
action in the standard Newtonian way, by subtracting kinetic energy.
The main result of our paper is the theorem stating that in the
stationary setting our model is equivalent to a pair of Weyl
equations. The crucial element of the proof is the observation that
our Lagrangian admits a factorization.
\end{abstract}


\maketitle

\section{Main result}
\label{Main result}

Throughout this paper we work on a 3-manifold $M$ equipped with
local coordinates $x^\alpha$, $\alpha=1,2,3$, and prescribed positive
metric $g_{\alpha\beta}$ which does not depend on time.
We extend the Riemannian 3-manifold $M$ to
a Lorentzian 4-manifold $\mathbb{R}\times M$ by adding the time
coordinate $x^0\in\mathbb{R}$. The metric on $\mathbb{R}\times M$ is
defined as
\begin{equation}
\label{Lorentzian metric}
\mathbf{g}_{{\bm{\alpha}}{\bm{\beta}}}=
\begin{pmatrix}
-1&0\\
0&g_{\alpha\beta}
\end{pmatrix}.
\end{equation}
Here and further on we use \textbf{bold} type for extended
quantities. Say, the use of bold type in tensor indices
${\bm{\alpha}},{\bm{\beta}}$ appearing in the LHS of formula
(\ref{Lorentzian metric}) indicates that these
run through the values $0,1,2,3$,
whereas the use of normal type in tensor indices
$\alpha,\beta$ appearing in the RHS of formula
(\ref{Lorentzian metric}) indicates that these
run through the values $1,2,3$.

All constructions presented in the
paper are local so we do not make \emph{a priori} assumptions on the
geometric structure of $\{M,g\}$.

The Weyl equation is the massless Dirac equation. It is the accepted
mathematical model for a massless neutrino field. The dynamical
variable (unknown quantity) in the Weyl equation is a two-component
complex-valued spinor field $\xi$
which is a function of time $x^0\in\mathbb{R}$ and local coordinates
$x^\alpha$ on $M$. The explicit form of the Weyl equation is
\begin{equation}
\label{Weyl equation}
i(\pm\sigma^0{}_{\dot ab}\partial_0
+\sigma^\alpha{}_{\dot ab}\nabla_\alpha)\xi^b=0.
\end{equation}
Here the $\sigma$ are Pauli matrices, $\partial_0$ is the time derivative
and $\nabla_\alpha$ is the covariant spatial
derivative, see Appendix~\ref{Spinor notation} for details.
Summation in (\ref{Weyl equation}) is carried out over the tensor
index $\alpha=1,2,3$ as well as over the spinor index $b=1,2$.
The use of the partial derivative
$\partial_0=\partial/\partial x^0$ in equation
(\ref{Weyl equation})
is justified by the fact that the time coordinate $x^0$ is fixed and
we allow only changes of coordinates $(x^1,x^2,x^3)$ which do
not depend on~$x^0$.

We see that the Weyl equation (\ref{Weyl equation}) is a system of
two ($\dot a=\dot1,\dot2$) complex linear partial differential equations on
the 4-manifold $\mathbb{R}\times M$ for two complex unknowns
$\xi^b$, $b=1,2$. The two choices of sign in (\ref{Weyl equation})
give two versions of the Weyl equation which differ by time
reversal. Thus, we have a pair of Weyl equations.

We will be interested in spinor fields of the form
\begin{equation}
\label{stationary spinor field}
\xi(x^0,x^1,x^2,x^3)=
e^{-ip_0x^0}\eta(x^1,x^2,x^3)
\end{equation}
where
\begin{equation}
\label{varepsilon is nonzero}
p_0\ne0
\end{equation}
is a real number. Substituting
(\ref{stationary spinor field}) into (\ref{Weyl equation}) we get
the equation
\begin{equation}
\label{stationary Weyl equation}
\pm p_0\sigma^0{}_{\dot ab}\eta^b
+i\sigma^\alpha{}_{\dot ab}\nabla_\alpha\eta^b=0
\end{equation}
which we shall call \emph{the stationary Weyl equation}. The
difference between equations
(\ref{Weyl equation}) and (\ref{stationary Weyl equation})
is that
in equation (\ref{Weyl equation}) the spinor field $\xi$ ``lives''
on  the Lorentzian 4-manifold $\mathbb{R}\times M$
whereas
in equation (\ref{stationary Weyl equation}) the spinor field $\eta$
``lives'' on  the Riemannian 3-manifold $M$.
Thus, the stationary Weyl equation is the Weyl equation with time
separated out. The relationship between
equations
(\ref{Weyl equation}) and (\ref{stationary Weyl equation})
is the same as between the wave equation and the Helmholtz equation.

The stationary Weyl equation (\ref{stationary Weyl equation}) is the
object of study of this paper. We separated out time to
simplify the problem while retaining most of its essential features. Note
also that this separation of variables has a clear physical meaning:
the real number $p_0$ appearing in
(\ref{stationary spinor field})
and
(\ref{stationary Weyl equation})
is quantum mechanical energy.

The aim of our paper is to show that the stationary Weyl equation
(\ref{stationary Weyl equation}) can be reformulated in an
alternative (but mathematically equivalent) way using instead of a
spinor field a different set of dynamical variables. Namely, we view
our 3-manifold $M$ as an elastic continuum whose material points can
experience no displacements, only rotations, with rotations of
different material points being totally independent. The idea of
rotating material points may seem exotic, however it has long been
accepted in continuum mechanics within the Cosserat theory of
elasticity \cite{Co}. This idea also lies at the heart of the theory
of \emph{teleparallelism} (=~absolute parallelism
=~fernparallelismus), a subject promoted by A.~Einstein and
\'E.~Cartan \cite{MR543192,unzicker-2005-,MR2276051}. See Section
\ref{Discussion} for more details.

Rotations of material points of the 3-dimensional elastic continuum
are described mathematically by attaching to each geometric point of
the manifold $M$ an orthonormal basis, which gives a field of
orthonormal bases called the \emph{frame} or \emph{coframe},
depending on whether one prefers dealing with vectors or covectors.
Our mathematical model will be built on the basis of exterior
calculus (no need for covariant derivatives) so for us it will be
more natural to use the coframe.

The coframe $\vartheta$ is a triple of orthonormal covector fields
$\vartheta^j$, $j=1,2,3$, on the 3-manifold $M$. Each covector field
$\vartheta^j$ can be written more explicitly as
$\vartheta^j{}_\alpha$ where the tensor index $\alpha=1,2,3$
enumerates the components. The orthonormality condition for the
coframe can be represented as a single tensor identity
\begin{equation}
\label{constraint for coframe}
g=\delta_{jk}\vartheta^j\otimes\vartheta^k
\end{equation}
where $\delta_{jk}$ is the Kronecker delta.
For the sake of clarity we repeat formula~(\ref{constraint for coframe})
giving tensor indices explicitly and performing summation over frame
indices explicitly:
\[
g_{\alpha\beta}=\delta_{jk}\vartheta^j{}_\alpha\vartheta^k{}_\beta
=\vartheta^1{}_\alpha\vartheta^1{}_\beta
+\vartheta^2{}_\alpha\vartheta^2{}_\beta
+\vartheta^3{}_\alpha\vartheta^3{}_\beta
\]
where $\alpha$ and $\beta$ run through the values $1,2,3$.
We view the identity (\ref{constraint for coframe}) as a kinematic
constraint: the metric $g$ is given (prescribed) and the coframe
elements $\vartheta^j$ are chosen so that they satisfy
(\ref{constraint for coframe}), which leaves us with three real
degrees of freedom at every point of $M$.

As dynamical variables in our model we choose the coframe
$\vartheta$ and a positive density $\rho$. Our coframe and density
are functions of local coordinates $(x^1,x^2,x^3)$ on $M$ as well as of
time $x^0$.

At a physical level, making the density $\rho$ a
dynamical variable means that we view our continuum more like a
fluid rather than a solid. In other words, we allow the material to
redistribute itself so that it finds its equilibrium distribution.

Note that the total number of real dynamical degrees of freedom
contained in the coframe $\vartheta$ and positive density $\rho$
is four, exactly as in a two-component complex-valued spinor field $\xi$.
Moreover, it is known
(see Appendix \ref{Correspondence between coframes and spinors})
that a coframe $\vartheta$ and a (positive) density $\rho$ are
geometrically equivalent to a nonvanishing spinor field $\xi$ modulo
the sign of $\xi$.

The crucial element in our construction is the choice of potential
energy. It is known that in the purely rotational setting the
potential energy of a physically linear elastic continuum contains
three quadratic terms, with three real parameters (elastic moduli)
as factors. The three quadratic terms in potential energy correspond
to the three irreducible pieces of torsion. It is not \emph{a
priori} clear what the elastic moduli of ``world aether'' are. We
choose a potential energy which feels only one piece of torsion,
axial, and is, moreover, conformally invariant, i.e.~which is
invariant under the rescaling of the 3-dimensional metric $g$ by an
arbitrary positive scalar function. This leaves us with a unique,
up to rescaling by a positive constant, formula
(\ref{formula for potential energy})
for potential energy.

After the potential energy is chosen the remainder of our
construction is straightforward. We incorporate time into our action
in the standard Newtonian way, by subtracting kinetic energy. This
gives us the Lagrangian density (\ref{our Lagrangian density}).
As we are interested in
comparing our mathematical model with the Weyl equation,
we perform a change of dynamical variables and switch
from coframe $\vartheta$ and density $\rho$ to a spinor field $\xi$.
Our Lagrangian density now
takes the form (\ref{our Lagrangian density in terms of spinor}).
We write down the field equation (Euler--Lagrange
equation) for our Lagrangian density and observe that time
separates out if we seek stationary solutions
(\ref{stationary spinor field}); this separation of variables is
highly nontrivial because our field equation is nonlinear.
After separation of variables our Lagrangian density
takes the stationary form
(\ref{our Lagrangian density stationary}).

The main result of our paper is the following

\begin{theorem}
\label{main theorem}
A nonvanishing time-independent spinor field $\eta$
is a solution of the field equation
for our stationary Lagrangian density~(\ref{our Lagrangian density stationary})
if and only if it is a solution of one of
the two stationary Weyl equations~(\ref{stationary Weyl equation}).
\end{theorem}

Theorem \ref{main theorem} provides an elementary, in terms of
Newtonian mechanics and elasticity theory, interpretation of the
stationary Weyl equation. This interpretation is geometrically much
simpler than the traditional one as the mathematical description of
our model does not require the use of spinors, Pauli matrices or
covariant differentiation.

The only technical assumption contained in the statement of Theorem
\ref{main theorem} and its proof is that the density does not vanish
which is equivalent to the spinor field not vanishing. At the moment
we do not know how to drop this technical assumption. We can only
remark that generically one would not expect a spinor field $\eta$
``living'' on a 3-manifold to vanish
as this would mean satisfying four real equations
$\operatorname{Re}\eta^1=\operatorname{Im}\eta^1=
\operatorname{Re}\eta^2=\operatorname{Im}\eta^2=0$
having at our disposal only three real variables $x^\alpha$, $\alpha=1,2,3$.

The crucial element of the proof of Theorem \ref{main theorem} is
the observation that our Lagrangian density admits factorization,
see formula (\ref{factorization formula}).
Thus, our argument is similar to the original argument
of Dirac, the difference being that we factorize the Lagrangian
whereas Dirac factorized the field equation (Klein--Gordon
equation). In our model factorizing the field equation is
impossible because the equation is nonlinear.

Our paper has the following structure.
In Section~\ref{Our model} we describe our mathematical model,
in Section~\ref{Switching to the language of spinors} we switch to
the language of spinors,
in Section~\ref{Separating out time} we separate out time
and in Section~\ref{Factorization of our Lagrangian} we
factorize our Lagrangian.
Section~\ref{Proof of main theorem} contains the proof
of Theorem~\ref{main theorem}.
In Section~\ref{Plane wave solutions} we analyze plane wave solutions
and in Section~\ref{Relativistic representation of our Lagrangian}
we give a relativistic representation of our Lagrangian.
The concluding discussion is presented in Section~\ref{Discussion}.

Our paper also has five appendices dealing with
issues of a more technical nature: Appendix~\ref{General notation}
describes our general notation, Appendix~\ref{Spinor notation}
describes our spinor notation,
Appendix~\ref{Correspondence between coframes and spinors}
gives the correspondence between coframes and spinors,
Appendix~\ref{Spinor representation of axial torsion and angular velocity}
gives the spinor representation of axial torsion and angular velocity
and Appendix~\ref{Toy model}
deals with a toy model which illustrates how a particular class of
nonlinear second order differential equations reduces to pairs
of linear first order equations.

\section{Our model}
\label{Our model}

In this section we describe in detail our mathematical model. At a
basic level it was already sketched out in Section~\ref{Main result}.

We need to write down the potential energy of a deformed Cosserat
continuum. The natural measure of deformations
caused by rotations of material points is the torsion
tensor defined
by the explicit formula
\begin{equation}
\label{definition of torsion}
T:=
\delta_{jk}\vartheta^j\otimes d\vartheta^k
\end{equation}
where $\,d\,$ denotes the exterior derivative.
Here ``torsion'' means ``torsion of the teleparallel connection''
with ``teleparallel connection'' defined by the condition that
the covariant derivative of each coframe element $\vartheta^j$ is
zero; see Appendix A of \cite{jmp2009} for a concise exposition.

Our construction of
potential energy follows the logic of classical linear elasticity
\cite{MR884707}, the only difference being that instead of a rank 2
tensor (strain) we deal with a rank 3 tensor (torsion). The logic of
classical linear elasticity dictates that we must first decompose
our measure of deformation (torsion) into irreducible pieces, with
irreducibility understood in terms of invariance under changes of
local coordinates preserving the metric $g_{\alpha\beta}$ at a given
point $P\in M$. It is known \cite{MR2419830} that torsion has three irreducible
pieces labeled by the adjectives \emph{axial}, \emph{vector} and
\emph{tensor}.
(Vector torsion is sometimes called trace torsion.)
The general formula for the potential energy of a
homogeneous isotropic linear elastic material contains squares of all
irreducible pieces with some constant coefficients in front. Thus,
the general formula for potential
energy should contain three free parameters (elastic moduli).

We, however, choose to construct our potential energy using only one piece of
torsion, namely, the axial piece given by the explicit formula
\begin{equation}
\label{definition of axial torsion}
T^\mathrm{ax}:=
\frac13\delta_{jk}\vartheta^j\wedge d\vartheta^k.
\end{equation}
Comparing (\ref{definition of axial torsion}) with (\ref{definition of torsion})
we see that axial torsion has a very simple meaning: it is the
totally antisymmetric part of the torsion tensor ($T$ is
antisymmetric only in the last pair of indices whereas
$T^\mathrm{ax}$ is antisymmetric in all three). In other words,
$T^\mathrm{ax}$ is a 3-form.

We chose the axial piece of torsion because it has two remarkable
properties.
\begin{itemize}
\item
The definition of axial torsion
(\ref{definition of axial torsion}) is very simple in
that it does not involve the metric. In a sense, axial torsion
(3-form) is an analogue of the electromagnetic field tensor (2-form)
from Maxwell's theory.
\item
Axial torsion possesses the property of
conformal covariance, i.e.~scales nicely under conformal rescalings
of the metric.
Indeed, it is easy to see that if we rescale our coframe as
\begin{equation}
\label{rescaling coframe}
\vartheta^j\mapsto e^h\vartheta^j
\end{equation}
where $h:M\to\mathbb{R}$ is an arbitrary scalar function,
then our metric scales as
\begin{equation}
\label{rescaling metric}
g_{\alpha\beta}\mapsto e^{2h}g_{\alpha\beta}
\end{equation}
and axial torsion scales as
\begin{equation}
\label{rescaling axial torsion}
T^\mathrm{ax}\mapsto e^{2h}T^\mathrm{ax}
\end{equation}
without the derivatives of $h$ appearing. The fact that axial
torsion is conformally covariant was previously observed by
Yu.~N.~Obukhov~\cite{MR664159} and J.~M.~Nester~\cite{MR1149210}.
\end{itemize}

We take the potential energy of our continuum to be
\begin{equation}
\label{formula for potential energy}
P(x^0):=\int_M\|T^\mathrm{ax}\|^2\rho\,dx^1dx^2dx^3.
\end{equation}
It is easy to see that the potential energy
(\ref{formula for potential energy})
is conformally
invariant: it does not change if we rescale our coframe
as (\ref{rescaling coframe}) and
our density as
\begin{equation}
\label{rescaling density}
\rho\mapsto e^{2h}\rho.
\end{equation}
This follows from formulas
(\ref{rescaling axial torsion}), (\ref{rescaling metric})
and
\[
\|T^\mathrm{ax}\|^2=\frac1{3!}
T^\mathrm{ax}_{\alpha\beta\gamma}T^\mathrm{ax}_{\kappa\lambda\mu}
g^{\alpha\kappa}g^{\beta\lambda}g^{\gamma\mu}.
\]

We take the kinetic energy of our continuum to be
\begin{equation}
\label{formula for kinetic energy}
K(x^0):=\int_M\|\dot\vartheta\|^2\rho\,dx^1dx^2dx^3
\end{equation}
where $\dot\vartheta$ is the 2-form
\begin{equation}
\label{definition of dot vartheta}
\dot\vartheta:=
\frac13\delta_{jk}\vartheta^j\wedge\partial_0\vartheta^k
\end{equation}
(compare with (\ref{definition of axial torsion})).
The 2-form (\ref{definition of dot vartheta}) can, of course, be written as
\begin{equation}
\label{dot vartheta via angular velocity}
\dot\vartheta=\frac23*\omega
\end{equation}
where
\begin{equation}
\label{definition of angular velocity}
\omega:=\frac12*(\delta_{jk}\vartheta^j\wedge\partial_0\vartheta^k)
\end{equation}
is the (pseudo)vector of angular velocity.
Hence,
(\ref{formula for kinetic energy}) is the standard expression for
the kinetic energy of a homogeneous isotropic Cosserat continuum. In
writing formula (\ref{formula for kinetic energy}) we assumed
homogeneity (properties of the material are the same at all points
of the manifold $M$) and isotropy (properties of the material are
invariant under rotations of the local coordinate system). We think
of each material point as a uniform ball possessing a moment of
inertia and without a preferred axis of rotation.

We now combine the potential energy (\ref{formula for potential energy}) and
kinetic energy (\ref{formula for kinetic energy}) to form the action
(variational functional) of our dynamic problem:
\begin{multline}
\label{our action}
S(\vartheta,\rho):=\int_{\mathbb{R}}(P(x^0)-K(x^0))\,dx^0
\\
=\int_{{\mathbb{R}}\times M}L(\vartheta,\rho)\,dx^0dx^1dx^2dx^3
\end{multline}
where
\begin{equation}
\label{our Lagrangian density}
L(\vartheta,\rho):=(\|T^\mathrm{ax}\|^2-\|\dot\vartheta\|^2)\rho
\end{equation}
is our Lagrangian density. Note that our construction of the action
(\ref{our action}) out of potential and kinetic energies is
Newtonian (compare with classical elasticity or even the harmonic
oscillator in classical mechanics).

Our field equations (Euler--Lagrange equations) are obtained by
varying the action (\ref{our action}) with respect to the
coframe $\vartheta$ and density $\rho$. Varying with respect to the
density $\rho$ is easy: this gives the field equation
$\|T^\mathrm{ax}\|^2=\|\dot\vartheta\|^2$ which is equivalent to
$L(\vartheta,\rho)=0$. Varying with respect to the coframe
$\vartheta$ is more difficult because we have to maintain the kinematic
constraint (\ref{constraint for coframe}); recall that the metric is
assumed to be prescribed (fixed).

A technique for varying the coframe with kinematic constraint
(\ref{constraint for coframe}) was described in Appendix B of
\cite{jmp2009}.
We, however, do not write down the field equations
for the Lagrangian density (\ref{our Lagrangian density}) explicitly.
We note only that they are highly nonlinear and do not appear to
bear any resemblance to the linear Weyl equation
(\ref{Weyl equation}).

\begin{remark}
\label{rigid rotations of the coframe}
The 3-form $T^\mathrm{ax}$ and 2-form $\dot\vartheta$
are invariant under rigid rotations of the coframe, i.e.~under special
orthogonal transformations (\ref{orthogonal transformation}) with constant
$O^j{}_k$. Hence, our Lagrangian density (\ref{our Lagrangian density})
is invariant under rigid rotations of
the coframe and, accordingly, solutions of our field equations
whose coframes differ by a rigid rotation
can be collected into equivalence classes.
Further on we view coframes differing
by a rigid rotation as equivalent.
\end{remark}

\section{Switching to the language of spinors}
\label{Switching to the language of spinors}

As pointed out in the previous section, varying the
coframe subject to the kinematic constraint
(\ref{constraint for coframe}) is not an easy task.
This technical difficulty can be overcome by switching
to a different dynamical variable. Namely, it is known, see Appendix
\ref{Correspondence between coframes and spinors}, that
in dimension~$3$
a coframe $\vartheta$ and a (positive) density $\rho$ are equivalent
to a nonvanishing spinor field $\xi$ modulo the sign of $\xi$. The
great advantage of switching to a spinor field $\xi$ is that there are no
kinematic constraints on its components, so the derivation
of field equations becomes absolutely straightforward.

We now need to substitute formulas
(\ref{formula for scalar}),
(\ref{formula for coframe elements 1 and 2}) and
(\ref{formula for coframe element 3})
into
(\ref{definition of axial torsion})
and
(\ref{definition of dot vartheta})
to get explicit expressions for
$T^\mathrm{ax}$ and $\dot\vartheta$ in terms of the spinor field
$\xi$.
The results are presented in Appendix
\ref{Spinor representation of axial torsion and angular velocity}.
Namely, formula
(\ref{axial torsion via spinor}) gives the spinor representation
of the 3-form $T^\mathrm{ax}$
whereas formulas~(\ref{angular velocity via spinor}) and
(\ref{dot vartheta via angular velocity}) give the spinor representation
of the 2-form $\dot\vartheta$.
We also know the spinor representation for our density $\rho$,
see formulas (\ref{formula for scalar}) and (\ref{formula for density}).
Substituting all these into formula
(\ref{our Lagrangian density})
we arrive at the following self-contained explicit spinor representation
of our Lagrangian density
\begin{multline}
\label{our Lagrangian density in terms of spinor}
L(\xi)=
\frac4{9\bar\xi^{\dot c}\sigma_{0\dot cd}\xi^d}
\Bigl(
[i(
\bar\xi^{\dot a}\sigma^\alpha{}_{\dot ab}\nabla_\alpha\xi^b
-
\xi^b\sigma^\alpha{}_{\dot ab}\nabla_\alpha\bar\xi^{\dot a}
)]^2
\\
-
\|i(
\bar\xi^{\dot a}\sigma_{\alpha\dot ab}\partial_0\xi^b
-
\xi^b\sigma_{\alpha\dot ab}\partial_0\bar\xi^{\dot a}
)\|^2
\Bigr)\sqrt{\operatorname{det}g}\,.
\end{multline}
Here and further on we write our Lagrangian density and our action
as $L(\xi)$ and $S(\xi)$ rather than $L(\vartheta,\rho)$ and $S(\vartheta,\rho)$,
thus indicating that we have switched to spinors. The nonvanishing
spinor field $\xi$ is the new dynamical variable and it will be
varied without any constraints.

Straightforward calculations show that the
field equation
for our Lagrangian
density~(\ref{our Lagrangian density in terms of spinor}) is
\begin{multline}
\label{field equation}
-\frac{4i}3\bigl(
(*T^\mathrm{ax})\sigma^\alpha{}_{\dot ab}\nabla_\alpha\xi^b
+\sigma^\alpha{}_{\dot ab}\nabla_\alpha((*T^\mathrm{ax})\xi^b)
\bigr)
\\
-\frac{8i}9\bigl(
\omega_\alpha\sigma^\alpha{}_{\dot ab}\partial_0\xi^b
+\sigma^\alpha{}_{\dot ab}\partial_0(\omega_\alpha\xi^b)
\bigr)
\\
-\rho^{-1}L\sigma_{0\dot ab}\xi^b
=0
\end{multline}
where the geometric quantities $*T^\mathrm{ax}$, $\omega$, $\rho$ and $L$
are expressed via the spinor field $\xi$ in accordance with formulas
(\ref{axial torsion via spinor}),
(\ref{angular velocity via spinor}),
(\ref{formula for scalar}),
(\ref{formula for density}) and
(\ref{our Lagrangian density in terms of spinor}).
The LHS of equation~(\ref{field equation}) is the spinor field
$F_{\dot a}$ appearing in the formula for the variation of the
action (\ref{our action}):
\[
\delta S
=\int_{{\mathbb{R}}\times M}
(F_{\dot a}\delta\bar\xi^{\dot a}+\bar F_a\delta\xi^a)
\sqrt{\operatorname{det}g}\ dx^0dx^1dx^2dx^3.
\]

We shall refer to equation (\ref{field equation})
as the \emph{dynamic} field equation, with ``dynamic'' indicating that
it contains the time derivative $\partial_0$.

\section{Separating out time}
\label{Separating out time}

Our dynamic field equation (\ref{field equation}) is highly nonlinear
and one does expect it to admit separation of variables.
Nevertheless,
we seek solutions of the form (\ref{stationary spinor field}).
Substituting formula
(\ref{stationary spinor field})
into formulas
(\ref{axial torsion via spinor}),
(\ref{angular velocity via spinor}),
(\ref{formula for scalar}),
(\ref{formula for density}) and
(\ref{our Lagrangian density in terms of spinor})
and using the identity (\ref{useful identity})
we get
\begin{equation}
\label{axial torsion via spinor stationary}
*T^\mathrm{ax}=-
\frac{
2i(
\bar\eta^{\dot a}\sigma^\alpha{}_{\dot ab}\nabla_\alpha\eta^b
-
\eta^b\sigma^\alpha{}_{\dot ab}\nabla_\alpha\bar\eta^{\dot a}
)
}
{
3\bar\eta^{\dot c}\sigma_{0\dot cd}\eta^d
}\,,
\end{equation}
\begin{equation}
\label{angular velocity via spinor stationary}
\omega_\alpha=
\frac{
2p_0
\bar\eta^{\dot a}\sigma_{\alpha\dot ab}\eta^b
}
{
\bar\eta^{\dot c}\sigma_{0\dot cd}\eta^d
}\,,
\end{equation}
\begin{equation}
\label{formula for density stationary}
\rho=\bar\eta^{\dot a}\sigma_{0\dot ab}\eta^b
\,\sqrt{\operatorname{det}g}\,,
\end{equation}
\begin{multline}
\label{our Lagrangian density in terms of spinor stationary}
L(\eta)=
\frac{16}{9\bar\eta^{\dot c}\sigma_{0\dot cd}\eta^d}
\Bigl(
\Bigl[\frac i2(
\bar\eta^{\dot a}\sigma^\alpha{}_{\dot ab}\nabla_\alpha\eta^b
-
\eta^b\sigma^\alpha{}_{\dot ab}\nabla_\alpha\bar\eta^{\dot a}
)\Bigr]^2
\\
-
(p_0\bar\eta^{\dot a}\sigma_{0\dot ab}\eta^b)^2
\Bigr)\sqrt{\operatorname{det}g}\,.
\end{multline}
Note that the geometric quantities
(\ref{axial torsion via spinor stationary})--(\ref{our Lagrangian density in terms of spinor stationary})
do not depend on time $x^0$, which simplifies the next step:
substituting
(\ref{stationary spinor field})
into our dynamic field equation
(\ref{field equation}),
using the identity (\ref{useful identity})
and dividing through by the common factor
$e^{-ip_0x^0}$
we get
\begin{multline}
\label{field equation stationary}
-\frac{4i}3\bigl(
(*T^\mathrm{ax})\sigma^\alpha{}_{\dot ab}\nabla_\alpha\eta^b
+\sigma^\alpha{}_{\dot ab}\nabla_\alpha((*T^\mathrm{ax})\eta^b)
\bigr)
\\
-\frac{32p_0^2}9
\sigma_{0\dot ab}\eta^b
-\rho^{-1}L\sigma_{0\dot ab}\eta^b
=0\,.
\end{multline}

The remarkable feature of formulas
(\ref{axial torsion via spinor stationary})--(\ref{field equation stationary})
is that they do not contain dependence on time $x^0$.
Thus, we have shown that our
dynamic field equation
(\ref{field equation})
admits separation of
variables, i.e.~one can seek solutions in the
form~(\ref{stationary spinor field}).

We shall refer to equation (\ref{field equation stationary})
as the \emph{stationary} field equation, with ``stationary'' indicating that
time $x^0$ has been separated out.

Consider now the action
\begin{equation}
\label{our action stationary}
S(\eta):=\int_ML(\eta)\,dx^1dx^2dx^3
\end{equation}
where $L(\eta)$ is our ``stationary'' Lagrangian density
(\ref{our Lagrangian density in terms of spinor stationary}).
It is easy to see that our stationary field equation
(\ref{field equation stationary}) is the Euler--Lagrange
equation for our
``stationary'' action~(\ref{our action stationary}).

In the remainder of the paper we do not use the explicit form of the
stationary field equation (\ref{field equation stationary}), dealing only
with the stationary Lagrangian density
(\ref{our Lagrangian density in terms of spinor stationary})
and the stationary action (\ref{our action stationary}).
We needed the explicit form of field equations,
dynamic and stationary, only to justify
separation of variables.

It appears that the underlying group-theoretic reason
for our nonlinear dynamic field equation
(\ref{field equation})
admitting separation
of variables is the fact that our model is $\mathrm{U}(1)$-invariant,
i.e.~it is invariant under the multiplication of the spinor field $\xi$
by a complex constant of modulus 1. Hence, it is feasible that
one could have performed the separation of variables argument
without even writing down the explicit form of field equations.

We give for reference a more compact representation of our stationary
Lagrangian density
(\ref{our Lagrangian density in terms of spinor stationary})
in terms of axial torsion $T^\mathrm{ax}$
(see formula (\ref{axial torsion via spinor stationary}))
and density~$\rho$
(see formula~(\ref{formula for density stationary})):
\begin{equation}
\label{our Lagrangian density stationary}
L(\eta)=\Bigl(
\|T^\mathrm{ax}\|^2-\frac{16}9p_0^2
\Bigr)\rho\,.
\end{equation}
Of course, formula (\ref{our Lagrangian density stationary})
is our original formula (\ref{our Lagrangian density})
with time separated out.
The choice of dynamical variables in the stationary Lagrangian density
(\ref{our Lagrangian density stationary}) is up to the user: one can either
use the time-independent spinor field~$\eta$ or, equivalently,
the corresponding time-independent coframe and time-independent density
(the latter are related to $\eta$ by formulas
(\ref{formula for scalar})--(\ref{formula for coframe element 3})
with $\xi$ replaced by $\eta$).
The important thing is that now our
dynamical variables are time-independent because we have separated out time.

The fact that we use the same notation $L$ both for the dynamic and
stationary Lagrangian densities should not cause problems
as in all subsequent sections, apart form Section
\ref{Relativistic representation of our Lagrangian},
we deal with the stationary case only.

\section{Factorization of our Lagrangian}
\label{Factorization of our Lagrangian}

Put
\begin{multline}
\label{Weyl Lagrangian density stationary}
L_{\pm}(\eta):=
\Bigl[
\frac i2(
\bar\eta^{\dot a}\sigma^\alpha{}_{\dot ab}\nabla_\alpha\eta^b
-
\eta^b\sigma^\alpha{}_{\dot ab}\nabla_\alpha\bar\eta^{\dot a}
)
\\
\pm p_0\bar\eta^{\dot a}\sigma^0{}_{\dot ab}\eta^b
\Bigr]\sqrt{\operatorname{det}g}\,.
\end{multline}
This is the Lagrangian density for the stationary Weyl equation
(\ref{stationary Weyl equation}).
Formula (\ref{Weyl Lagrangian density stationary})
can be written in more compact form as
\begin{equation}
\label{Weyl Lagrangian density stationary simplified}
L_\pm(\eta)=\Bigl(
-\frac34{*T^\mathrm{ax}}\mp p_0
\Bigr)\rho
\end{equation}
where $*T^\mathrm{ax}$ is the Hodge dual of axial torsion,
see formula~(\ref{axial torsion via spinor stationary}),
and $\rho$ is the density,
see formula (\ref{formula for density stationary}).
Comparing formulas
(\ref{our Lagrangian density stationary})
and
(\ref{Weyl Lagrangian density stationary simplified})
we get
\begin{equation}
\label{factorization formula}
L(\eta)=-\frac{32p_0}9
\frac{L_+(\eta)\,L_-(\eta)}{L_+(\eta)-L_-(\eta)}\,.
\end{equation}

Let us emphasize once again that throughout this paper we assume
that the density $\rho$ does not vanish, which is, of course, equivalent
to the spinor field not vanishing.
In view of formulas
(\ref{Weyl Lagrangian density stationary simplified})
and
(\ref{varepsilon is nonzero})
in the stationary case
the assumption $\rho\ne0$ can be
equivalently rewritten as
\begin{equation}
\label{density is nonzero}
L_+(\eta)\ne L_-(\eta)
\end{equation}
so the denominator in (\ref{factorization formula}) is nonzero.

Formula (\ref{factorization formula}) is the centerpiece of our paper:
it establishes the connection between Cosserat elasticity and the
Weyl equation. Moreover, the fact that the RHS of formula
(\ref{factorization formula}) contains a product of two Weyl
Lagrangian densities shows that we are essentially following Dirac's
factorization construction, the difference being that in the nonlinear
setting we cannot factorize equations and have to settle
for the next best thing --- factorizing the Lagrangian.

\section{Proof of Theorem \ref{main theorem}}
\label{Proof of main theorem}

Observe that the Lagrangian densities
$L_\pm$ defined by formula~(\ref{Weyl Lagrangian density stationary})
possess the property of scaling covariance:
\begin{equation}
\label{proof of theorem equation 1}
L_\pm(e^h\eta)=e^{2h}L_\pm(\eta)
\end{equation}
where $h:M\to\mathbb{R}$ is an arbitrary scalar function.
In fact, the Lagrangian density
of \emph{any} formally selfadjoint (symmetric) linear first order
partial differential operator
has the scaling covariance property (\ref{proof of theorem equation 1}).

We claim that the statement of Theorem \ref{main theorem}
follows from  formulas (\ref{factorization formula})
and
(\ref{proof of theorem equation 1}).
The proof presented below is an abstract one and does not depend
on the physical nature of the
dynamical variable $\eta$, the only requirement being that it is an element
of a vector space so that scaling makes sense.

Note that formulas
(\ref{factorization formula})
and
(\ref{proof of theorem equation 1})
imply that the Lagrangian density $L$
possesses the property of scaling covariance,
so all three of our Lagrangian densities,
$L$, $L_+$ and $L_-$,
have this property.
Note also that if $\eta$ is a solution of the field equation for
some Lagrangian density $\mathcal{L}\,$
possessing the property of scaling covariance
then $\mathcal{L}(\eta)=0$. Indeed, let us perform a scaling
variation of our dynamical variable
\begin{equation}
\label{scaling variation}
\eta\mapsto\eta+h\eta
\end{equation}
where $h:M\to\mathbb{R}$ is an arbitrary ``small'' scalar function
with compact support. Then
$0=\delta\!\int\!\mathcal{L}(\eta)=2\int h\mathcal{L}(\eta)$
which holds for arbitrary $h$ only if $\mathcal{L}(\eta)=0$.

In the remainder of the proof the variations of $\eta$ are arbitrary
and not necessarily of the scaling type (\ref{scaling variation}).

Suppose that $\eta$ is a solution of the field equation for
the Lagrangian density $L_+$.
[The case when $\eta$ is a solution of the field equation for
the Lagrangian density $L_-$ is handled similarly.]
Then $L_+(\eta)=0$ and, in view of formula (\ref{density is nonzero}),
$L_-(\eta)\ne0$.
Varying $\eta$ we get
\begin{multline*}
\delta\!\int\!L(\eta)
=
-\frac{32p_0}9\Bigl(
\int
\frac{L_-(\eta)}
{L_+(\eta)-L_-(\eta)}
\,\delta L_+(\eta)
\\
+
\int
L_+(\eta)
\,\delta\frac{L_-(\eta)}
{L_+(\eta)-L_-(\eta)}
\Bigr)
=\frac{32p_0}9\int\delta L_+(\eta)
\\
=\frac{32p_0}9\,\delta\!\int\!L_+(\eta)
\end{multline*}
so
\begin{equation}
\label{formula for variation of our action}
\delta\!\int\!L(\eta)=\frac{32p_0}9\,\delta\!\int\!L_+(\eta)\,.
\end{equation}
We assumed that $\eta$ is a solution of the field equation for
the Lagrangian density $L_+$ so
$\delta\!\int\!L_+(\eta)=0$ and
formula~(\ref{formula for variation of our action}) implies that
$\delta\!\int\!L(\eta)=0$. As the latter is true for an
arbitrary variation of $\eta$ this means that
$\eta$ is a solution of the field equation for the Lagrangian
density~$L$.

Suppose that $\eta$ is a solution of the field equation for
the Lagrangian density $L$.
Then $L(\eta)=0$ and formula~(\ref{factorization formula})
implies that either $L_+(\eta)=0$ or $L_-(\eta)=0$;
note that in view of (\ref{density is nonzero}) we cannot have simultaneously
$L_+(\eta)=0$ and $L_-(\eta)=0$.
Assume for definiteness that $L_+(\eta)=0$.
[The case when $L_-(\eta)=0$ is handled similarly.]
Varying~$\eta$ and repeating the argument from the previous paragraph
we arrive at (\ref{formula for variation of our action}).
We assumed that $\eta$ is a solution of the field equation for
the Lagrangian density $L$ so
$\delta\!\int\!L(\eta)=0$ and formula
(\ref{formula for variation of our action}) implies that
$\delta\!\int\!L_+(\eta)=0$. As the latter is true for an
arbitrary variation of $\eta$ this means that
$\eta$ is a solution of the field equation for the Lagrangian
density~$L_+$. $\square$

\section{Plane wave solutions}
\label{Plane wave solutions}

Suppose that $M=\mathbb{R}^3$ is Euclidean 3-space equipped
with Cartesian coordinates $x=(x^1,x^2,x^3)$
and standard Euclidean metric
(\ref{standard Euclidean metric}).
In this section we construct a special class of explicit solutions
of the field equations for our Lagrangian density
(\ref{our Lagrangian density}).
This construction is presented in the language of spinors.

Let us choose Pauli matrices
(\ref{Pauli matrices for standard Euclidean metric})
and seek solutions of the form
\begin{equation}
\label{Plane wave solutions equation 2}
\xi(x^0,x^1,x^2,x^3)=
e^{-i(p_0x^0+p\cdot x)}\zeta
\end{equation}
where $p_0$ is a real number as in
formulas
(\ref{stationary spinor field})
and
(\ref{varepsilon is nonzero}),
$p=(p_1,p_2,p_3)$ is a real constant covector
and $\zeta\ne0$ is a constant spinor.
We shall call solutions of the type
(\ref{Plane wave solutions equation 2}) \emph{plane wave\/}.
In seeking plane wave solutions
what we are doing is separating out all the variables,
namely, the time variable $x^0$ and the spatial variables
$x=(x^1,x^2,x^3)$.

Our dynamic field equation (\ref{field equation})
is highly nonlinear so it is
not \emph{a priori} clear that one can seek solutions in the form of plane
waves. However, plane wave solutions (\ref{Plane wave solutions equation 2})
are a special case of
stationary solutions
(\ref{stationary spinor field})
and these have already been analyzed in preceding sections.
Namely, Theorem~\ref{main theorem} gives us an algorithm for the calculation
of all plane wave solutions (\ref{Plane wave solutions equation 2})
by reducing the problem to a pair of stationary Weyl equations
(\ref{stationary Weyl equation})
for the time-independent spinor field
\begin{equation}
\label{Plane wave solutions new equation 2}
\eta(x^1,x^2,x^3)=e^{-ip\cdot x}\zeta .
\end{equation}
Substituting formulas
(\ref{definition of zeroth Pauli matrix}),
(\ref{Pauli matrices for standard Euclidean metric})
and
(\ref{Plane wave solutions new equation 2})
into equation (\ref{stationary Weyl equation}) we get
\begin{equation}
\label{Plane wave solutions equation 3}
\begin{pmatrix}
\mp p_0+p_3&p_1-ip_2\\
p_1+ip_2&\mp p_0-p_3
\end{pmatrix}
\begin{pmatrix}
\zeta^1\\
\zeta^2
\end{pmatrix}=0.
\end{equation}
The determinant of the matrix in the LHS of equation
(\ref{Plane wave solutions equation 3}) is
$p_0^2-p_1^2-p_2^2-p_3^2$ so this system has a nontrivial solution
$\zeta$ if and only if $p_0^2-p_1^2-p_2^2-p_3^2=0$. Our model
is invariant under rotations of the Cartesian coordinate system
(orthogonal transformations of the coordinate system preserving
orientation) so without loss of generality
we can assume that
\begin{equation}
\label{Plane wave solutions equation 4}
p_1=p_2=0,\qquad p_3=\pm p_0
\end{equation}
where the $\pm$ sign is chosen to agree with that in
equation~(\ref{Plane wave solutions equation 3}),
i.e.~upper sign in (\ref{Plane wave solutions equation 4})
corresponds to upper sign in (\ref{Plane wave solutions equation 3})
and same for lower signs.
Substituting formulas~(\ref{Plane wave solutions equation 4})
into equation
(\ref{Plane wave solutions equation 3})
and recalling our assumption~(\ref{varepsilon is nonzero})
we conclude that,
up to scaling by a nonzero complex factor,
we have
\begin{equation}
\label{Plane wave solutions equation 5}
\zeta^d=
\begin{pmatrix}
1\\
0
\end{pmatrix}.
\end{equation}

Combining formulas (\ref{Plane wave solutions equation 2}),
(\ref{Plane wave solutions equation 4}) and
(\ref{Plane wave solutions equation 5})
we conclude that for each
real $p_0\ne0$ our model admits, up to a rotation
of the coordinate system and complex scaling,
two plane wave solutions and that these plane wave solutions are given by the explicit
formula
\begin{equation}
\label{Plane wave solutions equation 6}
\xi^d=
\begin{pmatrix}
1\\
0
\end{pmatrix}e^{-ip_0(x^0\pm x^3)}.
\end{equation}

Let us now rewrite the plane wave solutions
(\ref{Plane wave solutions equation 6})
in terms of our original dynamical variables, coframe $\vartheta$
and density $\rho$. Substituting formulas
(\ref{definition of zeroth Pauli matrix}),
(\ref{Pauli matrices for standard Euclidean metric})
and
(\ref{Plane wave solutions equation 6})
into formulas
(\ref{formula for scalar})--(\ref{formula for coframe element 3})
we get $\rho=1$ and
\begin{multline}
\label{Plane wave solutions equation 7}
\vartheta^1{}_\alpha
=\begin{pmatrix}
\cos2p_0(x^0\pm x^3)\\
\sin2p_0(x^0\pm x^3)\\
0
\end{pmatrix},
\\
\vartheta^2{}_\alpha
=\begin{pmatrix}
-\sin2p_0(x^0\pm x^3)\\
\cos2p_0(x^0\pm x^3)\\
0
\end{pmatrix},
\quad
\vartheta^3{}_\alpha
=\begin{pmatrix}
0\\
0\\
1
\end{pmatrix}.
\end{multline}
Note that scaling the spinor $\zeta$ by a
nonzero complex factor is equivalent to scaling
the density $\rho$ by a positive real factor and
time shift $x^0\mapsto x^0+\operatorname{const}$.

We will now establish how many different (ones that cannot be
continuously transformed into one another) plane wave solutions we
have. To this end, we rewrite formula
(\ref{Plane wave solutions equation 7}) in the form
\begin{multline}
\label{Plane wave solutions equation 8}
\vartheta^1{}_\alpha
=\begin{pmatrix}
\cos2|p_0|(x^0+bx^3)\\
a\sin2|p_0|(x^0+bx^3)\\
0
\end{pmatrix},
\\
\vartheta^2{}_\alpha
=\begin{pmatrix}
-a\sin2|p_0|(x^0+bx^3)\\
\cos2|p_0|(x^0+bx^3)\\
0
\end{pmatrix},
\quad
\vartheta^3{}_\alpha
=\begin{pmatrix}
0\\
0\\
1
\end{pmatrix}
\end{multline}
where $a$ and $b$ can, independently, take values $\pm1$.
It may seem that we have a total of 4 different plane wave solutions.
Recall, however, that we can
perform rigid rotations of the coframe and that we have agreed
(see Remark~\ref{rigid rotations of the coframe} at the end
of Section \ref{Our model}) to view
coframes that differ by a rigid rotation as equivalent.
Let us perform a rotation of the coordinate system
\[
\begin{pmatrix}
x^1\\
x^2\\
x^3
\end{pmatrix}
\mapsto
\begin{pmatrix}
x^2\\
x^1\\
-x^3
\end{pmatrix}
\]
simultaneously with a rigid rotation of the coframe
\[
\begin{pmatrix}
\vartheta^1\\
\vartheta^2\\
\vartheta^3
\end{pmatrix}
\mapsto
\begin{pmatrix}
\vartheta^2\\
\vartheta^1\\
-\vartheta^3
\end{pmatrix}.
\]
It is easy to see that the above transformations
turn a solution of the form
(\ref{Plane wave solutions equation 8})
into a solution of this form again only with
\[
a\mapsto-a,
\qquad
b\mapsto-b.
\]
Thus, the numbers $a$ and $b$ on their own do not characterize
different plane wave solutions. Different plane wave solutions are
characterized by the number $c:=ab$ which can take two values, $+1$
and $-1$.

We have established that for a given positive frequency $|p_0|$
we have two essentially different types of
plane wave solutions. These can be written, for example, as
\begin{multline}
\label{Plane wave solutions equation 9}
\vartheta^1{}_\alpha
=\begin{pmatrix}
\cos2|p_0|(x^0+x^3)\\
\pm\sin2|p_0|(x^0+x^3)\\
0
\end{pmatrix},
\\
\vartheta^2{}_\alpha
=\begin{pmatrix}
\mp\sin2|p_0|(x^0+x^3)\\
\cos2|p_0|(x^0+x^3)\\
0
\end{pmatrix},
\quad
\vartheta^3{}_\alpha
=\begin{pmatrix}
0\\
0\\
1
\end{pmatrix}.
\end{multline}

The plane wave solutions (\ref{Plane wave solutions equation 9})
describe traveling waves of rotations. Both waves travel with
the same velocity (speed of light) in the negative $x^3$-direction.
The difference between the two solutions is in the direction of rotation
of the coframe:
if we fix the spatial coordinate~$x^3$ and look at the evolution of
(\ref{Plane wave solutions equation 9}) as a function of time $x^0$
or if we fix time $x^0$ and look at the evolution of
(\ref{Plane wave solutions equation 9}) as a function of the spatial coordinate $x^3$
then one solution describes a clockwise rotation whereas the other solution
describes an anticlockwise rotation.
We identify one of the solutions
(\ref{Plane wave solutions equation 9})
with a left-handed massless neutrino and
the other with a right-handed massless antineutrino.

The bottom line is that our model gives the correct number, two, of
distinct plane wave solutions.

\section{Relativistic representation of our Lagrangian}
\label{Relativistic representation of our Lagrangian}

In this section we work on the 4-manifold
$\mathbb{R}\times M$ equipped with Lorentzian metric
(\ref{Lorentzian metric}). This manifold is an extension of the
original 3-manifold~$M$. We use \textbf{bold} type for extended quantities.

We extend our coframe as
\begin{equation}
\label{Relativistic equation 1}
{\bm{\vartheta}}{}^0{}_{\bm{\alpha}}=
\begin{pmatrix}1\\0_\alpha\end{pmatrix},
\end{equation}
\begin{equation}
\label{Relativistic equation 2}
{\bm{\vartheta}}{}^j{}_{\bm{\alpha}}=
\begin{pmatrix}0\\\vartheta^j{}_\alpha\end{pmatrix},
\qquad j=1,2,3,
\end{equation}
where the bold tensor index $\bm{\alpha}$ runs through the values 0,
1, 2, 3, whereas its non-bold counterpart $\alpha$ runs through the
values 1, 2, 3. In particular, the $0_\alpha$ in formula
(\ref{Relativistic equation 1}) stands for a column of three zeros.

Throughout this section
our original 3-dimensional coframe $\vartheta$ is allowed to depend
on time $x^0$ in an arbitrary (not necessarily harmonic) manner,
as long as the kinematic constraint
(\ref{constraint for coframe}) is maintained. Thus, our only
restriction on the choice of extended 4-dimensional coframe
$\bm{\vartheta}$ is formula (\ref{Relativistic equation 1}) which
says that the zeroth element of the coframe is prescribed as the
conormal to the original Riemannian 3-manifold $M$.

The extended metric (\ref{Lorentzian metric}) is expressed via the extended
coframe (\ref{Relativistic equation 1}) and (\ref{Relativistic equation 2}) as
\begin{equation}
\label{Relativistic equation 3}
\mathbf{g}=
\mathbf{o}_{\mathbf{j}\mathbf{k}}
{\bm{\vartheta}}{}^{\mathbf{j}}\otimes{\bm{\vartheta}}{}^{\mathbf{k}}
\end{equation}
where $\mathbf{o}_{\mathbf{j}\mathbf{k}}
=\mathbf{o}^{\mathbf{j}\mathbf{k}}:=\mathrm{diag}(-1,+1,+1,+1)$
(compare with formula (\ref{constraint for coframe})).
The extended axial torsion is
\begin{multline}
\label{Relativistic equation 4}
\mathbf{T}^\mathrm{ax}:=
\frac13\mathbf{o}_{\mathbf{j}\mathbf{k}}
{{\bm{\vartheta}}{}^{\mathbf{j}}\wedge d{\bm{\vartheta}}{}^{\mathbf{k}}}
\\
=\frac13
(-\underset{=0}
{
\underbrace{
{\bm{\vartheta}}{}^0\!\wedge d{\bm{\vartheta}}{}^0
}
}
\!+{{\bm{\vartheta}}{}^1\!\wedge d{\bm{\vartheta}}{}^1}
\!+{{\bm{\vartheta}}{}^2\!\wedge d{\bm{\vartheta}}{}^2}
\!+{{\bm{\vartheta}}{}^3\!\wedge d{\bm{\vartheta}}{}^3}
)
\end{multline}
where $\,d\,$ denotes the exterior derivative on $\mathbb{R}\times M$
(compare with formula (\ref{definition of axial torsion})).
Formula (\ref{Relativistic equation 4}) can be rewritten~as
\begin{equation}
\label{Relativistic equation 5}
\mathbf{T}^\mathrm{ax}
=T^\mathrm{ax}-{\bm{\vartheta}}{}^0\!\wedge\dot\vartheta
\end{equation}
with $T^\mathrm{ax}$ and $\dot\vartheta$ defined by formulas
(\ref{definition of axial torsion})
and
(\ref{definition of dot vartheta}) respectively.
Squaring (\ref{Relativistic equation 5}) we get
$\|\mathbf{T}^\mathrm{ax}\|^2=\|T^\mathrm{ax}\|^2-\|\dot\vartheta\|^2$
which implies that our Lagrangian density (\ref{our Lagrangian density})
can be rewritten as
\begin{equation}
\label{Relativistic equation 6}
L(\vartheta,\rho)=\|\mathbf{T}^\mathrm{ax}\|^2\rho\,.
\end{equation}

The point of the arguments presented in this section was to show that
if one adopts the relativistic point of view then our Lagrangian density
(\ref{our Lagrangian density}) takes the especially simple form
(\ref{Relativistic equation 6}).
Formula (\ref{Relativistic equation 6}) is also useful in
that it allows us to see that our Lagrangian density
is invariant under conformal
rescalings of the 4-dimensional Lorentzian metric $\mathbf{g}$:
the arguments from Section \ref{Our model}
(see formulas (\ref{rescaling coframe})--(\ref{rescaling axial torsion})
and (\ref{rescaling density}))
carry over to the 4-dimensional setting without change.

A consistent pursuit of the relativistic
approach would require the variation of all four elements of the extended
coframe, giving three extra dynamical degrees of freedom
(Lorentz boosts in three directions). We do not do this in the current paper,
assuming instead that
the zeroth element of the extended coframe is specified by
formula~(\ref{Relativistic equation 1}).

\section{Discussion}
\label{Discussion}

The mathematical model presented in Section \ref{Our model} is, effectively,
a special case of the theory of teleparallelism
\cite{MR543192,unzicker-2005-,MR2276051}.
Modern reviews of teleparallelism can be found in
\cite{MR2419830,MR583723,MR1661422,MR1617858,MR1871425,MR1976711}.
The differences between our mathematical model
and those commonly used in teleparallelism are as follows.
\begin{itemize}
\item
We assume the metric to be prescribed (fixed) whereas in
teleparallelism it is traditional to view the metric as a dynamical
variable. In other words, in teleparallelism it is
customary to view~(\ref{constraint for coframe}) not as a kinematic constraint
but as a definition of the metric and, consequently, to vary the
coframe without any constraints. This is not surprising as
most, if not all, authors who contributed to teleparallelism came to
the subject from General Relativity.
\item
We take the density of our continuum $\rho$ to be
a dynamical variable whereas in teleparallelism
the tradition is to prescribe it as
$\rho=\sqrt{\operatorname{det}g}\,$.
Taking $\rho$ to be a dynamical variable is, of course, equivalent
to introducing an extra real positive scalar field $\rho/\sqrt{\operatorname{det}g}$
into our model
\item
We choose a very particular Lagrangian density (\ref{Relativistic equation 6})
containing only one irreducible piece of torsion (axial) whereas in
teleparallelism it is traditional to choose a more general
Lagrangian containing all three pieces
(axial, vector and tensor): see formula (26)
in \cite{MR2419830}.
In choosing our
particular Lagrangian density (\ref{Relativistic equation 6})
we were guided by the principles of conformal invariance,
simplicity and analogy with Maxwell's theory.
\end{itemize}

The main result of our paper is Theorem \ref{main theorem}
which establishes that in the
stationary setting (prescribed harmonic oscillation in time)
our mathematical model is equivalent to a pair
of massless Weyl equations (\ref{Weyl equation}).
The advantage of our approach is that it makes the Weyl equation look
natural to someone with a continuum mechanics background. The downside
is that our mathematical model is nonlinear which makes it look unnatural
to someone with a quantum mechanical background.

The situation here has a certain similarity with integrable systems.
Say, the Korteweg–-de Vries equation
(mathematical model of waves on shallow water surfaces)
is nonlinear but the
inverse scattering transform reduces it
to the analysis of a spectral problem for a linear Sturm--Liouville operator.
In our paper we go the other way round, reformulating the
spectral problem for the linear Weyl operator as a nonlinear equation
from continuum mechanics.

From a purely mathematical viewpoint Theorem \ref{main theorem} is
unusual in that it states that a (particular) second order partial
differential equation is equivalent to a pair of first order partial
differential equations, which is actually hard to believe. Indeed,
let us choose a 2-dimensional hypersurface $S$ on the 3-manifold $M$
and set a Cauchy problem on this surface. When dealing with a second
order partial differential equation one expects to be able to prescribe
the value of the spinor field $\eta$ on the surface $S$
as well as its normal derivative, whereas when dealing with a first
order partial differential equation one expects to be able to prescribe
the value of the spinor field $\eta$ only (the value of
the normal derivative of $\eta$ on the surface $S$ will be
determined by the equation). This argument appears to show that
there is no way a second order partial differential equation can be
reduced to a pair of first order equations. However, our second order
partial differential equation happens to be degenerate and does not
admit the setting of a standard Cauchy problem. This degeneracy
manifests itself in the property of scaling covariance of our
stationary Lagrangian density~(\ref{our Lagrangian density stationary}),
see Section~\ref{Proof of main theorem} for details.
Scaling covariance implies that our stationary Lagrangian
density~(\ref{our Lagrangian density stationary}) vanishes on
solutions of the (second order) field equation which means that the
value of the spinor field $\eta$ on the surface $S$ and its normal
derivative cannot be chosen independently.
In order to allay fears that there is something inherently wrong with our
construction we provide
in Appendix~\ref{Toy model}
an elementary example showing by means of an explicit calculation that
a second order differential equation
with Lagrangian of the form
(\ref{factorization formula}) and (\ref{proof of theorem equation 1})
does indeed reduce to a pair of first order equations.

Our construction exhibits a certain similarity with the Riccati
equation. Recall that the Riccati equation is a nonlinear first
order differential equation which reduces to a linear second order
differential equation. We go the other way round, reducing a
nonlinear second order equation to a pair of linear first order
equations. However, unlike the Riccati equation, our construction
works not only for ordinary differential equations but also for
partial differential equations.

Theorem \ref{main theorem} leaves us with two issues unresolved.
\begin{itemize}
\item[A]
What can be said about the general
case, when the spinor field $\xi$ is an arbitrary function
of all spacetime coordinates $(x^0,x^1,x^2,x^3)$
and is not necessarily of the form (\ref{stationary spinor field})?
\item[B]
What can be said about the relativistic version of our model
described in Section \ref{Relativistic representation of our Lagrangian}?
\end{itemize}
The two issues are, of course, related: both arise because in formulating
our basic model in Section \ref{Our model} we adopted the Newtonian approach
which specifies the time coordinate $x^0$ (``absolute time'').

We plan to tackle issue A by means of perturbation theory. Namely,
assuming the metric to be flat (as in Section \ref{Plane wave solutions}),
we start with a plane wave (\ref{Plane wave solutions equation 2})
and then seek the unknown spinor field $\xi$ in the form
\begin{equation}
\label{Discussion equation 1}
\xi(x^0,x^1,x^2,x^3)=
e^{-i(p_0x^0+p\cdot x)}\zeta(x^0,x^1,x^2,x^3)
\end{equation}
where $\zeta$ is a slowly varying spinor field. Here ``slowly
varying'' means that second derivatives of $\zeta$ can be
neglected compared to the first. Our conjecture is that the application
of a formal perturbation argument will yield the Weyl equation
(\ref{Weyl equation})
for the spinor field $\xi$.

We plan to tackle issue B by means of perturbation theory as well.
The relativistic version of our model has three extra field equations
corresponding to the three extra dynamical degrees of freedom
(Lorentz boosts in three directions). Our conjecture is that if we take
a solution of the nonrelativistic problem which is a perturbation of a plane
wave (as in the previous paragraph) then, at a perturbative level, this solution
will automatically satisfy the three extra field equations.
In other words, we conjecture
that our nonrelativistic model possesses
relativistic invariance at the perturbative level.

The detailed analysis of the two issues flagged up above will be the
subject of a separate paper.

\appendix

\section{General notation}
\label{General notation}

Our general notation mostly follows
\cite{jmp2009,prd2007},
the only major difference being that we changed
the signature of Lorentzian metric
$\,\mathbf{g}_{{\bm{\alpha}}{\bm{\beta}}}\,$
from $\,+---\,$ to $\,-+++\,$.
The latter is more natural when
promoting the Newtonian continuum mechanics approach.

We use Greek letters for tensor (holonomic) indices and Latin
letters for frame (anholonomic) indices.

We identify differential forms with covariant antisymmetric tensors.
Given a pair of real covariant antisymmetric tensors $P$ and $Q$ of
rank $r$ we define their dot product as
$
P\cdot Q:=\frac1{r!}P_{\alpha_1\ldots\alpha_r}Q_{\beta_1\ldots\beta_r}
g^{\alpha_1\beta_1}\ldots g^{\alpha_r\beta_r}
$.
We also define $\|P\|^2:=P\cdot P$.

All our constructions are local and occur in a neighborhood of a
given point $P$ of the 3-manifold $M$. We allow only changes of
local coordinates $x^\alpha$, $\alpha=1,2,3$, which preserve
orientation.

Working in local coordinates with specified orientation allows us to
define the Hodge star: we define the action of $\,*\,$ on a rank
$r$ antisymmetric tensor $R$ as
\begin{equation}
\label{definition of Hodge star}
(*R)_{\alpha_{r+1}\ldots\alpha_3}:=(r!)^{-1}\,\sqrt{\det g}\,
R^{\alpha_1\ldots\alpha_r}\varepsilon_{\alpha_1\ldots\alpha_3}
\end{equation}
where $\varepsilon$ is the totally antisymmetric quantity,
$\varepsilon_{123}:=+1$.

Coframes $\vartheta$ fall into two separate categories, depending
on the sign of
$\det\vartheta^j{}_\alpha$.
We choose to work with coframes satisfying the condition
\begin{equation}
\label{conditions on the coframe}
\det\vartheta^j{}_\alpha>0.
\end{equation}
Condition (\ref{conditions on the coframe}) means that orientation
encoded in our coframe agrees with that
encoded in our coordinate system.

An orthogonal transformation of a coframe is a linear map
\begin{equation}
\label{orthogonal transformation}
\vartheta^j\mapsto\tilde\vartheta^j=O^j{}_k\vartheta^k
\end{equation}
where the $O^j{}_k$ are real scalar functions
satisfying the condition
$\delta_{ji}\,O^j{}_k\,O^i{}_r=\delta_{kr}$.
Of course, orthogonal transformations map coframes into
coframes, i.e.~they preserve the kinematic constraint
(\ref{constraint for coframe}).
We call an orthogonal transformation special (or a rotation) if
the $O^j{}_k$ satisfy the additional condition
$\det O^j{}_k=+1$.
Any two coframes satisfying condition
(\ref{conditions on the coframe}) are related by a
special orthogonal transformation (rotation).

\section{Spinor notation}
\label{Spinor notation}

Our spinor notation mostly follows \cite{MR2176749},
the difference being that we changed
the signature of Lorentzian metric.

We use two-component complex-valued spinors (Weyl spinors) whose
indices run through the values $1,2$ or $\dot1,\dot2$. Complex
conjugation makes the undotted indices dotted and vice versa.

Define the ``metric spinor''
\begin{equation}
\label{metric spinor}
\epsilon_{ab}=\epsilon_{\dot a\dot b}=
\epsilon^{ab}=\epsilon^{\dot a\dot b}=
\left(
\begin{array}{cc}
0&-1\\
1&0
\end{array}
\right)
\end{equation}
with the first index enumerating rows and the second enumerating
columns. We will be using the spinor (\ref{metric spinor}) for
lowering and raising spinor indices.

We define
\begin{equation}
\label{definition of zeroth Pauli matrix}
\sigma_{0\dot ab}\!=\!
\sigma_0{}^{\dot ab}\!=
\!\left(
\begin{array}{cc}
1&0\\
0&1
\end{array}
\right)\!,
\
\sigma^0{}_{\dot ab}\!=\!
\sigma^{0\dot ab}\!=\!
-\!\left(
\begin{array}{cc}
1&0\\
0&1
\end{array}
\right)\!.
\end{equation}
The spinor (\ref{definition of zeroth Pauli matrix}) can also be
used for raising and lowering spinor indices. This is a feature of
the nonrelativistic setting, when we have a specified time
coordinate $t=x^0$ and transformations of spatial local coordinates
$x^\alpha$, $\alpha=1,2,3$, do not involve time.

Let $\mathfrak{v}$ be the real vector space of trace-free Hermitian $2\times2$
matrices $\sigma_{\dot ab}\,$. Pauli matrices
$\sigma_{\alpha\dot ab}\,$, $\alpha=1,2,3$,
are a basis in $\mathfrak{v}$ satisfying
\begin{equation}
\label{defining relation for Pauli matrices}
\sigma_{\alpha\dot ab}\sigma_\beta{}^{\dot ac}
+
\sigma_{\beta\dot ab}\sigma_\alpha{}^{\dot ac}
=-2g_{\alpha\beta}\delta_b{}^c
\end{equation}
where
$\,\sigma_\beta{}^{\dot ac}:=
\epsilon^{\dot a\dot e}\sigma_{\beta\dot ed}\epsilon^{cd}\,$.
Note that formula (\ref{defining relation for Pauli matrices}) automatically
implies an analogous formula for the extended metric
(\ref{Lorentzian metric}):
\begin{equation}
\label{defining relation for Pauli matrices extended}
\sigma_{{\bm{\alpha}}\dot ab}\sigma_{\bm{\beta}}{}^{\dot ac}
+
\sigma_{{\bm{\beta}}\dot ab}\sigma_{\bm{\alpha}}{}^{\dot ac}
=-2\mathbf{g}_{{\bm{\alpha}}{\bm{\beta}}}\delta_b{}^c
\end{equation}
where the bold tensor indices ${\bm{\alpha}},{\bm{\beta}}$ run
through the values $0,1,2,3$.

Of course, our Pauli matrices $\sigma_\alpha$, $\alpha=1,2,3$,
are not uniquely defined: if
$\sigma_\alpha=\sigma_{\alpha\dot ab}$ are Pauli matrices then so
are the matrices $U^*\sigma_\alpha U$ where $U$ is an arbitrary
special ($\operatorname{det}U=1$) unitary matrix-function.
Note also that under coordinate
transformations our Pauli matrices $\sigma_{\alpha a\dot b}$ transform
as components of a covector: this is indicated by the Greek
subscript $\alpha$.

Let us mention a useful identity for Pauli matrices, very similar
to (\ref{defining relation for Pauli matrices extended}) but with
contraction over tensor indices instead of spinor ones:
\begin{equation}
\label{useful identity}
\sigma_{{\bm{\alpha}}\dot ab}\sigma^{\bm{\alpha}}{}_{\dot cd}
=-2\epsilon_{\dot a\dot c}\epsilon_{bd}\,.
\end{equation}

We define the covariant derivatives of spinor fields as
\[
\nabla_\mu\xi^a=\partial_\mu\xi^a+\Gamma^a{}_{\mu b}\xi^b,
\qquad
\nabla_\mu\xi_a=\partial_\mu\xi_a-\Gamma^b{}_{\mu a}\xi_b,
\]
\[
\nabla_\mu\eta^{\dot a}=\partial_\mu\eta^{\dot a}
+\bar\Gamma^{\dot a}{}_{\mu\dot b}\eta^{\dot b},
\qquad
\nabla_\mu\eta_{\dot a}=\partial_\mu\eta_{\dot a}
-\bar\Gamma^{\dot b}{}_{\mu\dot a}\eta_{\dot b},
\]
where
$\bar\Gamma^{\dot a}{}_{\mu\dot b}=\overline{\Gamma^a{}_{\mu b}}$
and $\mu$ runs through the values $1,2,3$.
The explicit formula for the spinor connection coefficients
$\Gamma^a{}_{\mu b}$ can be derived from the following two conditions:
\begin{equation}
\label{condition 1}
\nabla_\mu\epsilon_{ab}=0,
\end{equation}
\begin{equation}
\label{condition 2}
\nabla_\mu\sigma^\alpha{}_{\dot ab}=0,
\end{equation}
where
\[
\nabla_\mu\sigma^\alpha{}_{\dot ab}=
\partial_\mu\sigma^\alpha{}_{\dot ab}
+\Gamma^\alpha{}_{\mu\beta}\sigma^\beta{}_{\dot ab}
-\bar\Gamma^{\dot c}{}_{\mu\dot a}\sigma^\alpha{}_{\dot cb}
-\Gamma^d{}_{\mu b}\sigma^\alpha{}_{\dot ad}
\]
and
$
\Gamma^\beta{}_{\alpha\gamma}=
\left\{{{\beta}\atop{\alpha\gamma}}\right\}:=
\frac12g^{\beta\delta}
(\partial_\alpha g_{\gamma\delta}
+\partial_\gamma g_{\alpha\delta}
-\partial_\delta g_{\alpha\gamma})
$
are the Christoffel symbols.
Conditions (\ref{condition 1}), (\ref{condition 2})
give an overdetermined system of linear algebraic equations for
$\mathrm{Re}\,\Gamma^a{}_{\mu b}$, $\mathrm{Im}\,\Gamma^a{}_{\mu b}$
the unique solution of which is
\begin{equation}
\label{spinor connection coefficient}
\Gamma^a{}_{\mu b}=-\frac14
\sigma_\alpha{}^{\dot ca}
\left(
\partial_\mu\sigma^\alpha{}_{\dot cb}
+\Gamma^\alpha{}_{\mu\beta}\sigma^{\beta}{}_{\dot cb}
\right).
\end{equation}
Observe that the sign in the RHS of formula
(\ref{spinor connection coefficient}) is different from that
of formula (A.9) in \cite{MR2176749}.
This is because we changed the signature of Lorentzian metric.

Note that for the standard Euclidean metric
\begin{equation}
\label{standard Euclidean metric}
g_{\alpha\beta}=\operatorname{diag}(1,1,1)
\end{equation}
the traditional choice of Pauli matrices is
\begin{equation}
\label{Pauli matrices for standard Euclidean metric}
\!
\sigma_{1\dot ab}=\!
\begin{pmatrix}
0&1\\
1&0
\end{pmatrix}\!,
\
\sigma_{2\dot ab}=\!
\begin{pmatrix}
0&-i\\
i&0
\end{pmatrix}\!,
\
\sigma_{3\dot ab}=\!
\begin{pmatrix}
1&0\\
0&-1\end{pmatrix}\!.
\end{equation}

\section{Correspondence between coframes and spinors}
\label{Correspondence between coframes and spinors}

In dimension $3$
a coframe $\vartheta$ and a (positive) density $\rho$ are equivalent
to a nonvanishing spinor field $\xi$ modulo the sign of $\xi$ in
accordance with the formulas
\begin{equation}
\label{formula for scalar}
s=\bar\xi^{\dot a}\sigma_{0\dot ab}\xi^b,
\end{equation}
\begin{equation}
\label{formula for density}
\rho=s\sqrt{\operatorname{det}g}\,,
\end{equation}
\begin{equation}
\label{formula for coframe elements 1 and 2}
(\vartheta^1+i\vartheta^2)_\alpha=s^{-1}
\epsilon^{\dot c\dot b}\sigma_{0\dot ba}\xi^a\sigma_{\alpha\dot cd}\xi^d,
\end{equation}
\begin{equation}
\label{formula for coframe element 3}
\vartheta^3{}_\alpha=s^{-1}
\bar\xi^{\dot a}\sigma_{\alpha\dot ab}\xi^b.
\end{equation}
The above formulas are a special case of those from \cite{MR0332092}.


We assume that our Pauli matrices are chosen in such a way that
the coframe $\vartheta$ defined by formulas
(\ref{formula for scalar}),
(\ref{formula for coframe elements 1 and 2}) and
(\ref{formula for coframe element 3})
satisfies condition (\ref{conditions on the coframe})
for all $\xi\ne0$.
Of course, the sign of $\operatorname{det}\vartheta^j{}_\alpha$ can
always be changed by switching from original Pauli matrices to their
complex conjugates.

Note that if we have the standard Euclidean
metric~(\ref{standard Euclidean metric}),
use traditional Pauli matrices
(\ref{Pauli matrices for standard Euclidean metric})
and take
\begin{equation}
\label{unit spinor}
\xi^a=
\begin{pmatrix}
1\\
0
\end{pmatrix}
\end{equation}
then formulas
(\ref{formula for scalar}),
(\ref{formula for coframe elements 1 and 2}) and
(\ref{formula for coframe element 3})
give us
\begin{equation}
\label{standard coframe}
\vartheta^j{}_\alpha=\delta^j{}_\alpha\,.
\end{equation}

\section{Spinor representation of axial torsion and angular velocity}
\label{Spinor representation of axial torsion and angular velocity}

We show in this appendix that
the Hodge dual of axial
torsion
(\ref{definition of axial torsion})
is expressed via the spinor field $\xi$ as
\begin{equation}
\label{axial torsion via spinor}
*T^\mathrm{ax}=-
\frac{
2i(
\bar\xi^{\dot a}\sigma^\alpha{}_{\dot ab}\nabla_\alpha\xi^b
-
\xi^b\sigma^\alpha{}_{\dot ab}\nabla_\alpha\bar\xi^{\dot a}
)
}
{
3\bar\xi^{\dot c}\sigma_{0\dot cd}\xi^d
}
\end{equation}
and that the vector of angular velocity $\omega$
defined by formula
(\ref{definition of angular velocity})
is expressed via the spinor field $\xi$ as
\begin{equation}
\label{angular velocity via spinor}
\omega_\alpha=
\frac{
i(
\bar\xi^{\dot a}\sigma_{\alpha\dot ab}\partial_0\xi^b
-
\xi^b\sigma_{\alpha\dot ab}\partial_0\bar\xi^{\dot a}
)
}
{
\bar\xi^{\dot c}\sigma_{0\dot cd}\xi^d
}\,.
\end{equation}
Note that formulas
(\ref{axial torsion via spinor})
and
(\ref{angular velocity via spinor})
are invariant under the rescaling of our spinor field by an arbitrary
nonvanishing real scalar function.

Formulas (\ref{axial torsion via spinor})
and
(\ref{angular velocity via spinor})
are proved by direct substitution of
formulas
(\ref{formula for scalar}),
(\ref{formula for coframe elements 1 and 2}) and
(\ref{formula for coframe element 3})
into
(\ref{definition of axial torsion})
and
(\ref{definition of angular velocity}) respectively.
In order to simplify calculations we observe
that the expressions in the left- and right-hand sides of formulas
(\ref{axial torsion via spinor})
and
(\ref{angular velocity via spinor})
have an invariant nature, hence it is sufficient to prove these formulas
for standard Euclidean metric~(\ref{standard Euclidean metric}),
traditional Pauli matrices
(\ref{Pauli matrices for standard Euclidean metric})
and at a point at which the spinor field takes
the value~(\ref{unit spinor}).

We have
\[
\xi^a=
\begin{pmatrix}
1+\delta\xi^1\\
\delta\xi^2
\end{pmatrix},
\]
\[
(\vartheta^1+i\vartheta^2)_\alpha=
\begin{pmatrix}
1+\delta\xi^1-\delta\bar\xi^{\dot1}\\
i+i\delta\xi^1-i\delta\bar\xi^{\dot1}\\
-2\delta\xi^2
\end{pmatrix},
\]
\[
\vartheta^3{}_\alpha=
\begin{pmatrix}
\delta\xi^2+\delta\bar\xi^{\dot2}\\
-i\delta\xi^2+i\delta\bar\xi^{\dot2}\\
1
\end{pmatrix},
\]
\begin{equation}
\label{curl of vartheta1 plus ivartheta2}
[\operatorname{curl}(\vartheta^1+i\vartheta^2)]_\alpha\!\!=\!\!
\begin{pmatrix}
-2\nabla_2\xi^2-\nabla_3(i\xi^1-i\bar\xi^{\dot1})\\
2\nabla_1\xi^2+\nabla_3(\xi^1-\bar\xi^{\dot1})\\
\nabla_1(i\xi^1-i\bar\xi^{\dot1})-\nabla_2(\xi^1-\bar\xi^{\dot1})
\end{pmatrix}\!\!,
\end{equation}
\begin{equation}
\label{curl of vartheta3}
[\operatorname{curl}\vartheta^3]_\alpha=
\begin{pmatrix}
-\nabla_3(-i\xi^2+i\bar\xi^{\dot2})\\
\nabla_3(\xi^2+\bar\xi^{\dot2})\\
\nabla_1(-i\xi^2+i\bar\xi^{\dot2})-\nabla_2(\xi^2+\bar\xi^{\dot2})
\end{pmatrix},
\end{equation}
\begin{equation}
\label{time derivative of vartheta1 plus ivartheta2}
[\partial_0(\vartheta^1+i\vartheta^2)]_\alpha=
\begin{pmatrix}
\partial_0\xi^1-\partial_0\bar\xi^{\dot1}\\
i\partial_0\xi^1-i\partial_0\bar\xi^{\dot1}\\
-2\partial_0\xi^2
\end{pmatrix},
\end{equation}
\begin{equation}
\label{time derivative of vartheta3}
[\partial_0\vartheta^3]_\alpha=
\begin{pmatrix}
\partial_0\xi^2+\partial_0\bar\xi^{\dot2}\\
-i\partial_0\xi^2+i\partial_0\bar\xi^{\dot2}\\
0
\end{pmatrix}
\end{equation}
where $\operatorname{curl}u:=*du$.

We rewrite the formulas for $*T^\mathrm{ax}$ and $\omega$ in the form
\begin{multline}
\label{convenient formula for axial torsion}
*T^\mathrm{ax}
=\frac16(\vartheta^1-i\vartheta^2)
\cdot\operatorname{curl}(\vartheta^1+i\vartheta^2)
\\
+\frac16(\vartheta^1+i\vartheta^2)
\cdot\operatorname{curl}(\vartheta^1-i\vartheta^2)
+\frac13\vartheta^3\cdot\operatorname{curl}\vartheta^3,
\end{multline}
\begin{multline}
\label{convenient formula for angular velocity}
\omega
=\frac14(\vartheta^1-i\vartheta^2)
\times\partial_0(\vartheta^1+i\vartheta^2)
\\
+\frac14(\vartheta^1+i\vartheta^2)
\times\partial_0(\vartheta^1-i\vartheta^2)
+\frac12\vartheta^3\times\partial_0\vartheta^3
\end{multline}
where $u\cdot v:=u_\alpha v^\alpha$ (note the absence of complex
conjugation) and $u\times v:=*(u\wedge v)$.
Substituting formulas (\ref{standard coframe}),
(\ref{curl of vartheta1 plus ivartheta2})
and
(\ref{curl of vartheta3})
into formula (\ref{convenient formula for axial torsion})
we get
\[
*T^\mathrm{ax}
=-\frac{2i}3\Bigl[
\nabla_3\xi^1+(\nabla_1-i\nabla_2)\xi^2
-\nabla_3\bar\xi^{\dot1}-(\nabla_1+i\nabla_2)\bar\xi^{\dot2}\Bigr]
\]
which coincides with the RHS of formula
(\ref{axial torsion via spinor}).
Substituting formulas (\ref{standard coframe}),
(\ref{time derivative of vartheta1 plus ivartheta2})
and
(\ref{time derivative of vartheta3})
into formula (\ref{convenient formula for angular velocity})
we get
\[
\omega_\alpha
=i
\begin{pmatrix}
\partial_0\xi^2-\partial_0\bar\xi^{\dot2}\\
-i\partial_0\xi^2-i\partial_0\bar\xi^{\dot2}\\
\partial_0\xi^1-\partial_0\bar\xi^{\dot1}
\end{pmatrix}
\]
which coincides with the RHS of formula
(\ref{angular velocity via spinor}).

An alternative way of proving formulas of the type
(\ref{axial torsion via spinor})
and
(\ref{angular velocity via spinor})
is to choose Pauli matrices
$\sigma_{\bm{\alpha}}$, $\bm{\alpha}=0,1,2,3$,
in such a way that a given nonvanishing spinor field $\xi$ takes the
value (\ref{unit spinor}) in some neighborhood of a given point
(as opposed to only the point itself). This approach was adopted,
for example, in
\cite{jmp2009,MR787353,MR1030935,MR1038610}.

\section{Toy model}
\label{Toy model}

In this appendix we present a toy model showing that a second
order differential equation with Lagrangian of the form
(\ref{factorization formula}) and (\ref{proof of theorem equation 1})
reduces to a pair of first order equations.

We work on the real line $\mathbb{R}$ parametrized
by the coordinate $x$. The dynamical variable (unknown quantity) is
the scalar function $\eta:\mathbb{R}\to\mathbb{C}\setminus\{0\}$.
Differentiation in $x$ is denoted by $\nabla$.

Consider a pair of first order linear ordinary differential
equations
\begin{equation}
\label{Toy model equation 1}
i\nabla\eta\pm\eta=0.
\end{equation}
The  corresponding Lagrangians are
\begin{equation}
\label{Toy model equation 2}
L_{\pm}(\eta):=
\frac i2(
\bar\eta\nabla\eta
-
\eta\nabla\bar\eta
)
\pm|\eta|^2.
\end{equation}
Equations (\ref{Toy model equation 1})
are simplified versions of the stationary Weyl equations
(\ref{stationary Weyl equation})
and Lagrangians (\ref{Toy model equation 2})
are simplified versions of the stationary Weyl Lagrangians
(\ref{Weyl Lagrangian density stationary}). Note that the Lagrangians
(\ref{Toy model equation 2}) possess the property of scaling
covariance (\ref{proof of theorem equation 1})
where $h:\mathbb{R}\to\mathbb{R}$ is an arbitrary scalar function.

By analogy with (\ref{factorization formula}), put
\begin{equation}
\label{Toy model equation 3}
L(\eta)\!:=\!
\frac{2L_+(\eta)L_-(\eta)}{L_+(\eta)\!-\!L_-(\eta)}
\!=\!\left[\!
\frac{i(\bar\eta\nabla\eta\!-\!\eta\nabla\bar\eta)}{2|\eta|}
\!\right]^2\!-\!|\eta|^2.
\end{equation}
The corresponding field equation (Euler--Lagrange equation) is
\begin{multline}
\label{Toy model equation 4}
\!\!\!\!\!
i\!\left\{\!
\frac{(\nabla\eta)}{|\eta|}
\!-\!\frac{\eta(\bar\eta\nabla\eta\!-\!\eta\nabla\bar\eta)}{2|\eta|^3}
\!+\!\nabla\frac\eta{|\eta|}
\!\right\}\!\!
\left[
\!\frac{i(
\bar\eta\nabla\eta
\!-\!
\eta\nabla\bar\eta
)}{2|\eta|}
\!\right]
\\
-\eta=0
\end{multline}
where the last $\nabla$ in the curly brackets acts on all the terms
to the right, including those in the square brackets. Equation
(\ref{Toy model equation 4}) is a second order nonlinear
ordinary differential equation which does not appear to bear any
resemblance to the first order linear ordinary differential
equations (\ref{Toy model equation 1}).

Let us switch to the polar representation of the complex function $\eta\,$:
\begin{equation}
\label{Toy model equation 5}
\eta=re^{-i\varphi}
\end{equation}
where
$r:\mathbb{R}\to(0,+\infty)$ and
$\varphi:\mathbb{R}\to\mathbb{R}$
are the new dynamical variables (unknown quantities).
Substituting formula (\ref{Toy model equation 5})
into equation (\ref{Toy model equation 4}) and multiplying
by $e^{i\varphi}$ we arrive at the polar representation of our
field equation:
\[
2i(\nabla r)(\nabla\varphi)
+r(\nabla\varphi)^2
+ir\nabla\nabla\varphi
-r=0.
\]
Separating the real and imaginary parts we rewrite the latter as a
system or real equations
\[
r(\nabla\varphi)^2-r=0,
\qquad
2(\nabla r)(\nabla\varphi)
+r\nabla\nabla\varphi=0,
\]
which, in turn, is equivalent to
\begin{equation}
\label{Toy model equation 6}
\nabla\varphi=\mp1,
\qquad
\nabla r=0.
\end{equation}
This shows that a complex function $\eta$ is a solution of equation
(\ref{Toy model equation 4}) if and only if
it is a solution of one of the two equations
(\ref{Toy model equation 1}).

Of course, the explicit calculations carried out above were
unnecessary because the toy model considered in this appendix is
covered by the abstract argument presented in Section~\ref{Proof of main theorem}.
The point of these explicit calculations was to illustrate the degeneracy
of field equations for Lagrangians of the form
(\ref{factorization formula}) and (\ref{proof of theorem equation 1}):
looking at (\ref{Toy model equation 6}) one sees the
absence of second derivatives.

\begin{acknowledgments}
The authors are grateful to
C.~G.~B\"ohmer,
F.~E.~A.~Johnson and Yu.~N.~Obukhov for stimulating discussions.
\end{acknowledgments}

\bibliography{weyl12}

\end{document}